\documentclass[11pt]{article}
\usepackage{moriond,epsfig}

\def\be{\begin{equation}}
\def\ee{\end{equation}}
\def\bea{\begin{eqnarray}}
\def\eea{\end{eqnarray}}

\begin{document}
\vspace*{4cm}
\title{THE AIR-FLUORESCENCE YIELD}

\author{F. Arqueros, F. Blanco, D. Garcia-Pinto, M. Ortiz and J. Rosado}

\address{Departmento de Fisica Atomica, Molecular y Nuclear, Facultad de Ciencias Fisicas, \\Universidad Complutense de Madrid, Madrid 28040, Spain}

\maketitle\abstracts{Detection of the air-fluorescence radiation induced by the charged particles of extensive air showers is a
well-established technique for the study of ultra-high energy cosmic rays. Fluorescence telescopes provide a nearly calorimetric
measure of the primary energy. Presently the main source of systematic uncertainties comes from our limited accuracy in the
fluorescence yield, that is, the number of fluorescence photons emitted per unit of energy deposited in the atmosphere by the
shower particles. In this paper the current status of our knowledge on the fluorescence yield both experimental an theoretical
will be discussed.}

\section{Introduction}
Fluorescence telescopes have been successfully used for the detection of ultra-high energy cosmic rays ($> 10^{18}$ eV) since
the pioneering Fly's Eye experiment~\cite{flys_eye}. In this technique the fluorescence radiation induced by the charged
particles of the extensive air shower generated by a primary cosmic ray is registered at ground by wide-angle telescopes.
Assuming that the intensity of the air-fluorescence light is proportional to the energy deposited in the atmosphere by the
shower, this technique provides a nearly calorimetric measure of the energy of the primary cosmic ray. Therefore it has the
advantage, as compared with methods relying on simulations (e.g. surface arrays working in standalone mode), of being nearly
model independent. In spite of this advantage, fluorescence telescopes are presently limited by the uncertainty in the
fluorescence yield, that is, the calibration parameter which converts number of fluorescence photons into absolute energy units.
For instance in the Pierre Auger Observatory~\cite{auger} the uncertainty in the fluorescence yield contributes a 14\% to the
total systematic error in the energy calibration which is presently 22\%.

In order to improve the accuracy of this parameter, dedicated laboratory experiments~\cite{experiments} are carrying out precise
measurements of the air-fluorescence emission. In these experiments an electron beam excites air at certain pressure and
temperature conditions. A large set of experimental parameters are measured, not only the absolute value of the fluorescence
yield but also the spectral features of the fluorescence radiation and the dependence with atmospheric parameters (pressure,
temperature, humidity, etc.). On the other hand, progress on the theoretical understanding of the various processes leading to
the air-fluorescence emission is being carried out~\cite{blanco}.
\begin{figure}[t]
\centering \epsfig{width=1\textwidth, file=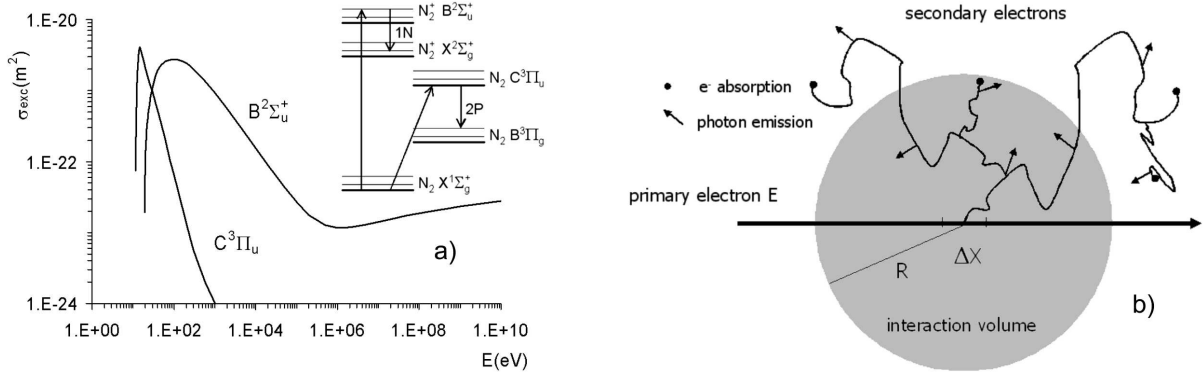} \caption{a) Molecular levels of N$_2$ and N$_2^+$ involved in the
generation of air-fluorescence and cross section versus electron energy for the excitation of the corresponding upper levels. b)
At high electron energy most of the fluorescence light is generated by secondary electrons.} \label{mol-levels}
\end{figure}
\section{The generation of air fluorescence excited by electrons}
\subsection{Physical processes}\label{subsec:fluorescence_proc}
Air-fluorescence in the near UV range (300 - 400 nm) is basically produced by the de-excitation of atmospheric nitrogen
molecules excited by the shower electrons. Most of the fluorescence light comes from the 2P System of N$_2$ and the 1N System of
N$_2^+$ (Fig. 1a). Excited molecules can also decay by collisions with other molecules (collisional quenching). This effect
which grows with pressure $P$, reduces the fluorescence intensity by a factor $1+P/P'_{\lambda}$. The characteristic pressure
$P'_{\lambda}$ is defined, for a given $v-v'$ band of wavelength $\lambda$, as the one for which collisional quenching and
radiative decay have the same probability.

Basically two different parameters are being used for the energy calibration of fluorescence telescopes. The first one
$\varepsilon_{\lambda}$ is the number of photons of a given molecular band emitted per electron and unit path length,
$\varepsilon_{\lambda} = N \times \sigma_{\lambda}/(1+P/P'_{\lambda})$, where $N$ is the density of nitrogen molecules and
$\sigma_{\lambda}$ is the cross section for the excitation of the molecular band. The second parameter is the \emph{fluorescence
yield} $Y_{\lambda}$, defined as the number of photons emitted per unit deposited energy.
\begin{equation}
\label{eq:FY} Y_{\lambda} = Y_{\lambda}^0 \frac{1}{1+P/P'_{\lambda}}\,,\qquad Y_{\lambda}^0 = \frac{A_{\lambda}}{({\rm d}E/{\rm
d}X)_{dep}}\,.
\end{equation}

$Y_{\lambda}^0$ is the fluorescence yield in the absence of quenching. $A_{\lambda}$ and $({\rm d}E/{\rm d}X)_{dep}$ are
respectively the number of emitted photons at zero pressure and the deposited energy both per unit mass thickness. The
fluorescence yield as defined in (1) is more useful for calorimetric applications. Notice that for the determination of
$Y_{\lambda}$, both photon number and deposited energy has to be measured in the same volume. This is particularly important for
laboratory experiments carried out in small gas chambers. In this case secondary electrons ejected in ionization processes might
escape the field of view of the optical system before depositing all the energy (Fig. 1b). In next section the role of secondary
electrons in the generation of air-fluorescence light is described.
\subsection{Secondary electrons}\label{subsec:sec_electr}
Secondary electrons from ionization processes are the main source of fluorescence light, since the excitation cross sections
show a fast decrease with energy (Fig. 1a), in particular the one for the 2P system. A high energy electron loses energy as a
result of collisions with air molecules. Ionization processes give rise to the ejection of secondary electrons which deposit
their energy within a certain distance from the interaction point (Fig. 1b). The average energy deposited per unit mass
thickness inside a given volume around the interaction point can be expressed as
\begin{equation}
\label{energy-dep} \rho \frac{{\rm d}E_{dep}}{{\rm d}X}= N_{air}\{<E_{dep}^0> + <E_{dep}>\} \sigma_{ion}(E)\,,
\quad\label{edep_0} <E_{dep}^0> = <E_{exc}>\frac{\sigma_{exc}}{\sigma_{ion}} + I + <E_{exc}^{ion}>\,,
\end{equation}
where $\rho$ is the air density, $N_{air}$ is the number of air molecules per unit volume and $\sigma_{ion}$ is the ionization
cross section. The average energy deposited in the medium by the primary electron per primary ionization process $<E_{dep}^0>$
is obtained from several molecular parameters~\footnote{ionization potential I, total excitation cross section $\sigma_{exc}$,
average excitation energy of neutral molecules $<E_{exc}>$ and of ionized molecules $<E_{exc}^{ion}>$.}. The energy deposited in
the volume by the secondary electrons $<E_{dep}>$ is calculated by a dedicated simulation~\cite{blanco}. Figure 2a) shows the
result for a sphere of radius $R$ (Fig. 1b). As expected, the deposited energy depends on $PR$ and for an unlimited medium, $PR
\rightarrow \infty$, equals the energy loss predicted by the Bethe-Bloch theory.

Neglecting the collisional quenching, the number of photons emitted per electron and per unit path length can be expressed by
$\varepsilon _{\lambda}(P) = \rho A_{\lambda} = N \{\sigma_{\lambda} (E) + \alpha_{\lambda}(E,P)\sigma_{ion}(E)\}$, where
$\alpha_{\lambda}(E,P)$ is the average number of photons generated inside the volume per secondary electron, also calculated in
the simulation. A very simple expression for $Y_{\lambda}^0$ can be obtained from the above equations
\begin{figure}[t]
\centering \epsfig{width=1\textwidth, file=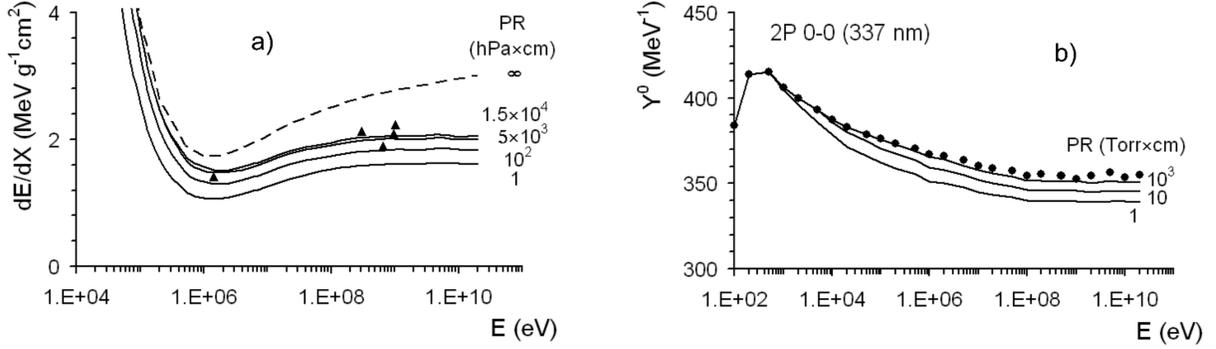} \caption{a) Continuous lines represent the energy deposited per unit
mass thickness versus electron energy for several values of $PR$. Dashed line is the total energy loss of the electron.
Triangles represent the relative values of the air-fluorescence yield measured by Kakimoto \textit{et al.} b) Fluorescence yield
at zero pressure versus primary energy for the 337 nm band.} \label{dep_energy}
\end{figure}

\begin{equation}
\label{Phi0} Y_{\lambda}^0 = \frac{N}{N_{air}}\times
\frac{\frac{\sigma_{\lambda}}{\sigma_{ion}}+\alpha_{\lambda}}{<E_{dep}^0>+<E_{dep}>}\,,
\end{equation}

This procedure allows theoretical predictions on the absolute value of $Y_{\lambda}^0$ and its dependence on the electron energy
as shown below.
\subsection{Fluorescence emission versus deposited energy}\label{subsec:dep_energy}
The energy calibration of fluorescence telescopes relies on the assumption that the intensity of fluorescence light is
proportional to the energy deposited in the atmosphere, that is, the fluorescence yield is assumed to be independent on the
electrons energy. The validity of this assumption can be theoretically checked by means of the model described above. Fig. 2b)
shows $Y^0$ versus $E$ for the most intense band of the 2P system (0-0 transition at 337 nm). The results shown in this plot can
be summarized as follows. The fluorescence yield decreases with $E$ about a 10\% in the range 1 keV - 1 MeV and about 4\% in the
interval 1 MeV - 20 GeV. This smooth dependence of the fluorescence yield on $E$ has no impact on the energy calibration of
fluorescence telescopes. The proportionality assumption has been also verified experimentally by several
groups~\cite{5th_FW_SP}.
\section{The dependence of the fluorescence yield on atmospheric parameters}
Fluorescence yield depends on pressure, temperature $T$ and humidity. Thus for a precise energy calibration of fluorescence
telescopes these dependencies have to be determined accurately.

As mentioned above collisional quenching reduces the fluorescence emission by a factor $1+P/P'_{\lambda}$. In the general case,
for a mixture of gases (e.g. nitrogen, oxygen, water vapor, etc.), the characteristic pressure obeys the law
\begin{equation}
\label{eq:1_P} \frac{1}{P'} = \sum_i{\frac{f_i}{P'_i}}\,,\qquad P'_i = \frac{kT}{\tau}\frac{1}{\sigma_{Ni}
\bar{v}_{Ni}}\,,\qquad \bar{v}_{Ni}=\sqrt{\frac{8kT}{\pi \mu_{Ni}}}\,,
\end{equation}
where $f_i$ is the fraction of molecules of type $i$ in the mixture, $\sigma_{Ni}$ is the collisional cross section which
depends on the particular band, and $v_{Ni}$ and $\mu_{Ni}$ are the relative velocity and reduced mass of the two body system
N-i respectively.

The experimental procedure for the determination of the dependence of fluorescence yield on the above parameters is the
following. At a fixed temperature the dependence of fluorescence intensity on pressure is measured for dry air. This measure, if
properly carried out~\footnote{the effect of secondary electrons escaping the field of view might introduce systematic errors.},
allows a determination of $P'$ and therefore the dependence of the fluorescence yield on pressure at a fixed temperature.
Experimental values of $P'$ for the molecular bands of the 2P and 1N systems in dry air at room temperature have been reported
by many authors~\cite{experiments}. The most complete set of $P'$ values have been reported very recently by AIRFLY~\cite{ave}
improving the accuracy of previous measurements. This set of values are being used by the Pierre Auger Observatory~\cite{auger}
for the calculation of the dependence of the fluorescence yield versus altitude~\footnote{the contribution of the pressure
dependence to the total uncertainty in the energy determination has been reduced to a 1\%.}.

The $P'$ parameter depends on temperature because the collision frequency grows with $\sqrt{T}$ as predicted by the kinetic
theory of gases. In addition the collisional cross section depends on the kinetic energy of the encounters following a power law
($\sim T^{\alpha}$). Assuming this effect is negligible, the temperature dependence of the fluorescence yield can be easily
predicted by equation (\ref{eq:1_P}). Recently some experimental works~\cite{5th_FW_SP} have found a noticeable variation of the
collisional cross section with temperature. According to the preliminary values reported by AIRFLY~\cite{5thFW_airfly_T},
neglecting this effect results in an overestimation of the fluorescence yield by an amount going up to $\approx$ 20\% for the 1N
(0-0) 391 nm band.

Water molecules have a significant cross section for the air-fluorescence quenching and therefore humidity modifies the value of
$P'$. Several authors~\cite{5th_FW_SP} have measured the dependence of fluorescence intensity on humidity. A decrease of the
fluorescence yield up to a 20\% is found (at 100\% relative humidity). From these measurements, values of the characteristic
pressure for the quenching with water molecules $P'_{H_2O}$ have been determined for the main molecular bands of nitrogen.
\begin{figure}[t]
\centering \epsfig{width=1\textwidth, file=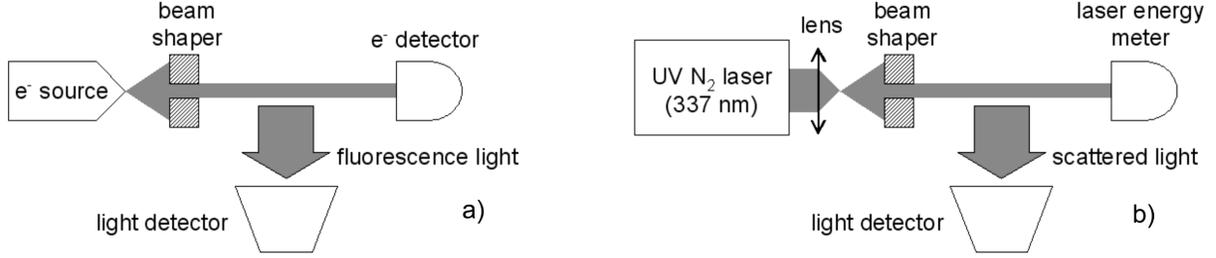} \caption{Comparison of fluorescence signal generated by a) an electron
beam with b) that from Rayleigh scattering of a nitrogen laser. This procedure allows the absolute calibration of the optical
system.} \label{Ray_cal}
\end{figure}
\section{Absolute value}\label{subsec:experimental_absolute}
The accurate measurement of the absolute value of the fluorescence yield is an experimental challenge. The value is obtained as
the ratio $Y_{\lambda} = N_{\lambda}/(N_{e}\times EDEP)$. For the measurement of the absolute number of fluorescence photons in
the wavelength interval of interest $N_{\lambda}$, the efficiency of the various elements of the collection and detection system
has to be known accurately. The number of electrons traversing the observation region $N_{e}$ has to be absolutely measured as
well. Finally the total energy $EDEP$ deposited in the volume from where the registered fluorescence was emitted has to be
determined (usually by means of a Monte Carlo simulation). In order to reduce systematic errors in the optical calibration (e.g.
PMT quantum efficiency, transmission of optical elements, geometrical factors, etc.) some techniques have been developed, based
on the comparison with well known physical processes like Cherenkov emission or Rayleigh scattering (Fig. 3).

Several measurements of $Y_{\lambda}$ are presently available~\cite{experiments}. Unfortunately the comparison is not simple
since some authors report the experimental result of $\varepsilon_{\lambda}$ (i.e. photons/m) while others provide $Y_{\lambda}$
(i.e. photons/MeV). In addition the spectral intervals of the various experiments use to be different. A detailed summary of the
available results can be found elsewhere~\cite{5th_FW_SP}. Here we will compare some representative experimental data (Tab. 1).
For this comparison, measured values of $\varepsilon_{\lambda}$ are converted into fluorescence yields using our results on
deposited energy. Notice that deposited energy is weakly dependent on the size of the region and therefore a rough estimate of
the equivalent $R$ value is sufficient. From these results the fluorescence yield $Y_{337}$ for the most intense band, 2P (0-0)
at 337 nm, is calculated using the experimental relative intensities reported by AIRFLY~\cite{ave}. Finally the $Y_{337}$ values
have been normalized to 293 K temperature and 1013 hPa pressure using equations (1) and (4). This procedure is appropriate for a
comparison of measurements with typical uncertainties of about 13\% or higher. Results are shown in last column of Tab. 1.

Firstly, the $\varepsilon_{\lambda}$ values of Kakimoto \textit{et al.} in the range 300-400 nm at several energies have been
superimposed in Fig. 2a) to the energy deposited at atmospheric pressure assuming an observation volume with $R$ ranging between
5 and 15 cm. The comparison of fluorescence intensity (photons/m) with deposited energy has allowed the determination of the
fluorescence yield (photons/MeV) in that wavelength interval.

The $\varepsilon_{337}$ value of 1.021 photons/m from Nagano \textit{et al.} has been combined with the deposited energy for
$R\approx$ 5 cm giving the corresponding $Y_{337}$ value. For the determination of the fluorescence yield, both MACFLY and FLASH
calculate the deposited energy from a MC simulation. For these experiments only the conversion for wavelength intervals as well
as minor $T$ and $P$ corrections were necessary. Finally AIRFLY reports a preliminary value of $Y_{337}$ determined from the
ratio of the absolute number of photons and the energy deposited according to a GEANT4 simulation.

\small
\begin{table*}[t]
\caption{Comparison of data on fluorescence yields. Experimental results are used to infer the value of the fluorescence yield
for the 337 nm band at $T$ = 293 K and $P$ = 1013 hPa (last column). See text for details.}

\centering
\begin{tabular}{|lcccc|c|c|}
\hline

Experiment  &  ~~~~ $\Delta \lambda$  ~~~~ &  T &   P  & ~experimental result~ & $I_{337}/I_{\Delta \lambda}$ & $Y_{337}$ \\
 &  nm &  [K] &   [hPa]  &    &  & [MeV]$^{-1}$\\

\hline \hline
Kakimoto \textit{et al.}     & 300 - 400     & 288 & 1013 & see text     &  0.278    & 5.4 \\
\hline
Nagano \textit{et al.}       & 337           & 293 & 1013 & 1.021 ph./m  &  1        & 5.5 \\
\hline
MACFLY              & 290 - 440     & 296 & 1013 & 17.6 ph./MeV &  0.261    & 4.6 \\
\hline
FLASH 07            & 300 - 420     & 304 & 1013 & 20.8 ph./MeV &  0.276    & 5.6 \\
\hline
AIRFLY (prelim.)    & 337           & 291 & 993  & 4.12 ph./MeV &    1      & 4.0 \\
\hline
\end{tabular}
\end{table*}
\normalsize

\section{Conclusions}
Our understanding on the processes leading to generation of air fluorescence has increased significantly in the last
years~\cite{5th_FW_SP}. The world-wide campaign for the experimental determination of the fluorescence yield has achieved
remarkable results, in particular in the measurement of the various dependencies with atmospheric parameters. The fundamental
assumption of proportionality between fluorescence intensity and deposited energy has been verified both theoretically and
experimentally.

In regard with the determination of the absolute value of the fluorescence yield new data are available. However the
interpretation of the results is not straightforward. A comparison using the procedure discussed here shows a general agreement
with typical differences of about 15\%. For a real improvement in the accuracy of fluorescence telescopes an uncertainty better
than 10\% in the fluorescence yield is necessary. Several experiments claim high accuracy, for instance, the reported
uncertainty of the FLASH experiment is of about 8\%. In addition the AIRFLY collaboration will publish soon a final absolute
value with an error below 10\%. A discussion on these and other high accuracy measurements have been presented
elsewhere~\cite{5th_FW_SP}. Discrepancies between these experiments go beyond the reported accuracies and therefore some
experimental effort is still necessary to clarify the situation.
\section*{Acknowledgments}
This work has been supported by the Spanish MEyC (Ref.: FPA2006-12184-C02-01), CONSOLIDER - INGENIO2010 program CPAN
(CSD2007-000042) and CM-UCM (Ref.: 910600).

\end{document}